\documentclass[12pt,a4paper]{article}
\usepackage{amsmath}
\usepackage{amsfonts}
\usepackage{amssymb}
\usepackage{textcomp}
\usepackage[dvips]{graphicx}
\usepackage{epsfig}
\usepackage{bm}
\usepackage{dcolumn}
\usepackage{amsfonts}
\usepackage{color}
\usepackage[left=2cm,right=2cm,top=2cm,bottom=2cm]{geometry}

\begin{document}
\title{\textbf{
Deviation from Slow-Roll Regime in the EGB Inflationary Models with $r\sim N_e^{-1}$}}

\author{Ekaterina~O.~Pozdeeva\footnote{E-mail address:
pozdeeva@www-hep.sinp.msu.ru}\vspace*{3mm} \\
\small  Skobeltsyn Institute of Nuclear Physics, Lomonosov Moscow State University,\\
\small Leninskie Gory 1, Moscow  119991,  Russia}

\date{ \ }

\maketitle
\begin{abstract}
We consider Einstein--Gauss--Bonnet (EGB) inflationary models using the effective potential approach.
We present evolution equations in the slow-roll regime using the effective potential and the tensor-to-scalar ratio.
The choice of the effective potential is related to an expression of the spectral index in terms of e-folding number $N_e$.
The satisfaction of the slow-roll regime is mostly related to the form of the tensor-to-scalar ratio $r$. The case of $r\sim1/N^2_e$
leads to a generalization of $\alpha$-attractors inflationary parameters to Einstein--Gauss--Bonnet gravity with exponential effective potential.
 Moreover, the cosmological attractors include models with $r\sim1/N_e$. And we check the satisfaction of the slow-roll regime  during inflation for models with  $r\sim1/N_e$.
\end{abstract}

\section{Introduction}
The  observations data~\cite{Akrami:2018odb} allow to check different types of inflationary models due to the known values of the spectral index $n_s$, the~amplitude $A_s$ of scalar perturbations and the restriction to the tensor-to-scalar ratio $r$. The~first model to historically satisfy current observation constrains~\cite{Akrami:2018odb} is the $R^2$ inflationary model~\cite{Starobinsky:1980te,Starobinsky:1982,Starobinsky:1983} with:
\begin{equation}
\label{prediction}
n_s=1-\frac{2}{N_{e}+N_0},\qquad r=\frac{12}{(N_{e}+N_0)^{2}}
\end{equation}
in leading order of inverse e-folding number $1/(N_e+N_0)$, where $N_0$ is a constant, $N_e$ is number of e-foldings.
The generalizations of $R^2$ inflationary scenario~\cite{Mijic:1986iv, Maeda1988} were introduced as cosmological attractors models~\cite{Kallosh:2013hoa,Galante:2014ifa,Roest:2013fha} which lead to spectrum \eqref{prediction} and
 at the same time allow two different relations between the tensor-to-scalar ratio and e-folding number: $r\sim(N_{e}+N_0)^{-2}$  and  $r\sim(N_{e}+N_0)^{-1}$. The~case of $r\sim(N_{e}+N_0)^{-2}$ belongs to \mbox{$\alpha$-attractor} models. The cosmological attractor models include inflationary scenarios inspired by particle physics~\cite{Barvinsky:1994hx,Bezrukov:2007ep,Barvinsky:2008ia,Bezrukov:2010jz,Bezrukov:2013fka,Rubio:2018ogq,Cervantes-Cota1995,Elizalde:2014xva,Elizalde:2015nya}. Multi-fields inflationary scenarios~\cite{Kaiser:2012ak,Greenwood:2012aj,Dubinin:2017irg} with scalar fields non-minimally coupled with Ricci scalar
allow  $\alpha$-attractors approximation~\cite{Dubinin:2017irq}.

The Einstein--Gauss--Bonnet gravity is inspired  by string  theory framework as a quantum correction to general relativity~\cite{Antoniadis:1993jc,Torii:1996yi,Kawai1998,Calcagni:2005im,Cartier:2001is,Hwang:2005hb,Tsujikawa:2006ph,Cognola:2006sp}. The~ construction-appropriate inflationary scenarios in the Einstein--Gauss--Bonnet gravity is an actively studied problem
~\cite{Pozdeeva:2020apf,Soda2008,Guo:2009uk,Guo:2010jr,Jiang:2013gza,Koh:2014bka,Koh:2016abf,Koh:2018qcy,vandeBruck:2015gjd,JoseMathew,vandeBruck:2016xvt,Nozari:2017rta,Armaleo:2017lgr,Chakraborty:2018scm,Yi:2018gse,Odintsov:2018zhw,Nojiri:2019dwl,Kleidis:2019ywv,Rashidi:2020wwg,Odintsov:2020sqy,Odintsov:2020zkl,Hikmawan:2015rze}.
The models with the Gauss--Bonnet term and the Ricci scalar multiplied by functions of the field can be considered such generalizations of the models with minimal coupling~\cite{vandeBruck:2015gjd,JoseMathew}.
The appropriate inflationary scenarios can be obtained both numerically and analytically.
The analytical studying of inflationary scenarios can be performed using the e-folding numbers $N_e$ presentation\cite{Mukhanov:2013tua}. The~model of the Einstein--Gauss--Bonnet gravity leading to the $\alpha$ attractor inflationary parameters was reconstructed in~\cite{Pozdeeva:2020shl} and studied in~\cite{Pozdeeva:2021iwc} using the presentation of inflationary scenarios in terms of the e-folding number $N_e$.
In~\cite{Guo:2010jr}, the models inspired by chaotic inflation~\cite{Leach:2002ar,Linde:1983gd} with monomial potential $V\sim \phi^n$ and an inverse function before the Gauss--Bonnet term $\xi\sim\phi^{-n}$ were studied in a slow-roll regime. However, the~case of $n=2$ leads to deviation from the slow-roll
regime before the end of inflation, as in~the case $n=4$, the value of the spectral index becomes sufficiently small to satisfy recently observed data~\cite{Akrami:2018odb}. Models constructed in~\cite{Guo:2010jr} lead to the tensor-to-scalar ratio of the form $r\sim 1/N_e$.
In~\cite{Pozdeeva:2020apf}, the correction to function $\xi(\phi)$ was introduced to obtain appropriate inflationary scenarios. However,
 a more complicate form of the tensor-to-scalar ratio $r$ was obtained. In~the present paper, models without the introduction  of $V(\phi)$ and $\xi(\phi)$  leading to inflationary parameters of cosmological attractors with  $r\sim(N_{e}+N_0)^{-1}$ are considered. In~our consideration, the effective potential formulated for  Einstein--Gauss--Bonnet gravity~\cite{Pozdeeva:2019agu} applicable near de Sitter solution~\cite{Vernov:2021hxo} or in slow-roll regime~\cite{Pozdeeva:2020apf,Pozdeeva:2021iwc} is used.

The paper is organized as follows. In~Section~\ref{sec2}, the action and the evolution equations in the slow-roll regime are briefly introduced and presented in terms of e-folding numbers using the effective potential formulation. In~Section~\ref{sec3}, the slow-roll parameters
 {are presented using the effective potential, the tensor-to-scalar ratio and nonminimal coupling function. The consideration  includes the case of
minimal coupling of field with Ricci scalar.} The expressions of the effective potential leading
to an appropriate spectral index are considered. Subsequently, the model with an exponential effective potential
with $r\sim(8r_0)/(N_{e}+N_0)$ is studied and the breaking of the slow-roll regime due
to the small value of $r_0$ is clearly demonstrated. In~Section~\ref{sec4}, the results of our~consideration formulate the conclusion.

\section{Slow-Roll Regime in EGB~Gravity}\label{sec2}

In this paper, the gravity model with a scalar field is considered, nonminimally coupled with both the Ricci curvature scalar and
the Gauss--Bonnet term, as described by the following action:
\begin{equation}
\label{action1}
S=\int d^4x\frac{\sqrt{-g}}{2}\left[F(\phi)R-g^{\mu\nu}\partial_\mu\phi\partial_\nu\phi-2V(\phi)-\xi(\phi){\cal G}\right],
\end{equation}
where the functions $F(\phi)$, $V(\phi)$, and~$\xi(\phi)$ are differentiable ones,  $R$ is the Ricci scalar and:
\begin{equation*}
\mathcal{G}=R_{\mu\nu\rho\sigma}R^{\mu\nu\rho\sigma}-4R_{\mu\nu}R^{\mu\nu}+R^2
\end{equation*}
is the Gauss--Bonnet term. We assume that $F(\phi)>0$ and $V(\phi)>0$ during~inflation.

The system of evolution equations in the spatially flat Friedmann--Lema\^{i}tre--Robert\-son--Walker metric with $ds^2=-dt^2+a^2(t)(dx^2+dy^2+dz^2)$
 was obtained and considered in slow-roll approximation~in~\cite{Guo:2010jr,vandeBruck:2015gjd}: $$\dot{\phi}^2\ll V,\,|\ddot{\phi}|\ll 3H|\dot{\phi}|,\,4|\dot{\xi}|H\ll F,\,|\ddot{\xi}|\ll|\dot{\xi}|H,\,|\ddot{F}|\ll H|\dot{F}|\ll H^2F,$$
where $H=\dot{a}/a$ is the Hubble parameter, $a(t)$ is the scale factor, dots denote the derivatives with respect to the cosmic time  $t$.
In~\cite{Pozdeeva:2020shl,Pozdeeva:2021iwc},  the slow-roll  approximation of evolution equations in terms of e-folding number $N_e$ was formulated using ${dA}/{dt}=-H{dA}/{dN_e}$.
The obtained equations can be presented  as follows:
\begin{equation}
H^2\simeq\frac{W}{3}, \quad \left({\phi}^\prime\right)^2\simeq 4WV^{\,\prime}_{eff}\label{prime(phi)}.
\end{equation}
where a prime denotes the derivative with respect to $N_e$,  $W\equiv V/F$ and the effective potential~\cite{Pozdeeva:2019agu}:
\begin{equation}
\label{Veff}
V_{eff}={}-\frac{F^2}{4V}+\frac{\xi}{3}.
\end{equation}

In our consideration, we use $N_e=-\ln(a/a_e)$ following the choice of notations in Ref.~\cite{Mukhanov:2013tua,Roest:2013fha,Pozdeeva:2020shl,Pozdeeva:2021iwc}.
 Note that in many papers~\cite{Dubinin:2017irg,Pozdeeva:2020apf,Guo:2010jr,vandeBruck:2015gjd,Odintsov:2020sqy},
 $N=-N_e$ is used as a new independent variable for evolution equations.
The slow-roll parameters as functions of $N_e$ can be presented such as
\begin{eqnarray}
&&\epsilon_1=\frac12\ln^\prime(W), \quad \zeta_1=-\ln^\prime(F), \quad  \delta_1=-\frac{4W}{3F}\xi^\prime,\\
&&\epsilon_{i+1}=-\ln^\prime(\epsilon_i),\qquad\zeta_{i+1}=-\ln^\prime(\zeta_i),\qquad\delta_{i+1}=-\ln^\prime(\delta_i).
\end{eqnarray}

In our consideration, we use expressions for the tensor-to-scalar ratio $r$ and the spectral index  of scalar perturbations $n_s$  obtained in~\cite{Pozdeeva:2021iwc}:
\begin{eqnarray}
  r&=&\frac{32WV^{\,\prime}_{eff}}{F},  \label{r_i}\\
  n_s&=&1+\frac{V^{\,\prime\prime}_{eff}}{V^{\,\prime}_{eff}}.  \label{nsN}
\end{eqnarray}

The model of EGB gravity leading to cosmological attractor inflationary parameters with $r\sim(N_e+N_0)^{-2}$ was assumed in~\cite{Pozdeeva:2020shl}.
The model was later studied in  detail and generalized to EGB gravity models with the field nonminimally coupled with Ricci scalar.
In the next section, we consider EGB gravity models leading to the $n_s$ coinciding with $n_s$ of cosmological attractors
with $r\sim(N_e+N_0)^{-1}$.

\section{Application}\label{sec3}
We try to reproduce the cosmological attractor inflationary parameters with $r\sim1/(N_e+N_0)$
in EGB gravity. Accordingly, with \eqref{nsN} and the expression of the spectral index with the second order correction~\cite{Pozdeeva:2021iwc}, we can write:
\begin{equation}\label{ind}
 \frac{V_{eff}^{\prime\prime}}{V^{\,\prime}_{eff}}=-\frac{2}{N_e+N_0}+\frac{C_2}{(N_e+N_0)^{2}}
\end{equation}

The satisfaction of the equation is possible by two~ways:
\begin{enumerate}
  \item If  $C_2\neq 0$ we can suppose $V_{eff}=C_{eff}\exp\left(-\frac{C_2}{N_e+N_0}\right)$;
  \item And if $C_2= 0$, we can suppose  $V_{eff}=\frac{C_{eff}}{N_e+N_0}$.
\end{enumerate}

In~\cite{Pozdeeva:2021iwc}, the exponential presentation of effective potential with second order tensor-to-scalar ratio $r\sim1/(N_e+N_0)^2$ was considered. After~the choice of tensor-to-scalar ratio, the form of the potential will be related  to the choice of nonminimal coupling \eqref{r_i} by
\begin{equation}
 W=\left(\frac{F\,r}{32{V^{\,\prime}_{eff}}}\right)
 \end{equation}

 At the same time, the function $\xi$  can be presented through the effective potential and the tensor-to-scalar ratio:
\begin{equation}
 \xi=3\,{V_{eff}}+{\frac {24\,V^{\,\prime}_{eff}}{r}}
 \end{equation}

 The dependence of slow-roll parameters $\epsilon_1$, $\zeta_1$, $\delta_1$ from $N_e+N_0$ related with the effective potential,  { the tensor-to-scalar ratio and the~nonminimal coupling function:}
\begin{eqnarray}
   \epsilon_1 &=& \frac12\left({\frac {r^\prime }{r}}
-\frac{{V^{\prime\prime}_{eff}} }{{
{V^{\,\prime}_{eff}}}}+{\frac {F^\prime }{F }}\right)\\
   \delta_1&=& \left({\frac {r^\prime }{r }}-\frac{r}{8} -{
\frac {V^{\prime\prime}_{eff}  }{{ V^{\,\prime}_{eff}}}}\right )=\left(2\epsilon_1+\zeta_1-\frac{r}{8}\right)\\
   \zeta_1&=& -\frac{F^\prime}{F}
 \end{eqnarray}

We suppose that at the end of inflation, the nonminimal coupling function tends to $1$. To~simplify the analysis of only dealing with Gauss--Bonnet coupling, we consider the case of constant coupling. In~this case, the~expressions for the slow-roll parameters $\epsilon_1$ and $\delta_1$ can be simplified as follows:
\begin{eqnarray}
   \epsilon_1 = \frac12\left({\frac {r^\prime }{r}}
-\frac{{V^{\prime\prime}_{eff}} }{{
{V^{\,\prime}_{eff}}}}\right),\quad   \delta_1= \left({\frac {r^\prime }{r }}-\frac{r}{8} -{
\frac {{ V^{\prime\prime}_{eff}}  }{{ V^{\,\prime}_{eff}}}}\right )=2\epsilon_1-\frac{r}{8},
 \end{eqnarray}
the slow-roll parameters $\zeta_{i}$ are absent. We assume that in the case of
\begin{equation} r = \frac{8r_0}{(N_e+N_0)} \label{rFirstrOrder}
 \end{equation}
 the upper values of parameter $r_0$ are rather small to save the  slow-roll regime during inflation. In~the next section, our~supposition will be estimated.

 The first step of the discrimination of models is checking the values first order slow-roll parameters during inflation.
 The second step of  {the} discrimination is the consideration of the second order slow-roll parameters.  {During}~the analysis, one should
 remember the appropriate values of the tensor-to-scalar  ratio. As such, we have two variants of effective potentials leading
 to the same spectral index in leading order of inverse e-folding number,
and thus consider the corresponding effective potentials in two different~subsections.
  \subsection{Power-Law Effective~Potential}
In this subsection, the power-law variant of effective potential is considered:
\begin{equation}
 V_{eff}=\frac{C_{eff}}{(N_e+N_0)}
 \end{equation}
which leads to the exact reproduction of the  spectral index without second order correction. At~the same time, the supposition $r\sim(N_e+N_0)^{-2}$
 leads to constant potential. The~first slow-roll, $\epsilon_1=0$, has no~end.

 The supposition \eqref{rFirstrOrder} leads to the following model:
\begin{equation}\label{modelpower}
    V=-{\frac {{r_0}\,(N_e+N_0)}{4{C_{eff}}}},\quad \xi={\frac {3{C_{eff}}\, \left( {r_0}-1 \right) }{{ r_0}\,(N_e+N_0)}},\quad\xi^\prime=-{\frac {3{C_{eff}}\, \left( {r_0}-1 \right) }{{r_0}\,{(N_e+N_0)}^{2}}}
 \end{equation}
  and slow-roll parameters:
\begin{eqnarray}
 \epsilon_1 &=&  \frac{1}{2\,{(N_e+N_0)}},\quad \epsilon_2= \frac{1}{{(N_e+N_0)}}\\
  \delta_1&=&-{\frac {{ r_0}-1}{(N_e+N_0)}},\quad \delta_2=\epsilon_2
\end{eqnarray}

We suppose that the exit from inflation is defined at $N_0=1/2$ at which $\epsilon_1(N_e=0)=1$. At~the   $N_0=1/2$, the second
slow-roll parameters $\epsilon_2$ and $\delta_2$ reach $2$ , thus the slow roll regime is infracted during inflation  and the slow-roll
approach is not applicable to the considered~model.

 \subsection{Exponential Effective~Potential}
 The case of  {an} exponential potential  {and}  $r\sim (N_e+N_0)^{-2}$ was considered in~\cite{Pozdeeva:2020shl,Vernov:2021hxo}.
 Now,  {we consider the case of the exponential potential and } $r\sim (N+N_0)^{-1}$  {:}
\begin{eqnarray}
   V_{eff}&=&C_{eff}\exp\left(-\frac{C_2}{N_e+N_0}\right).\label{ExpEffPot}
  \end{eqnarray}

The choice of effective potential \eqref{ExpEffPot} and  tensor-to-scalar ratio \eqref{rFirstrOrder} leads to the following model in terms of e-folding number:
\begin{equation}\label{modelExpEffectivePot}
V= \frac{{r_0}\,(N_e+N_0)}{4{{C_{eff}}{{C_2}}\left( {\exp\left(-{\frac{{C_2}}{N_e}}\right)} \right)}},\quad \xi=\frac{3\,{C_{eff}}\,{\exp\left(-{\frac {{C_2}}{N_e+N_0}}\right)} \left( {r_0}\,(N_e+N_0)+{C_2} \right) } {{{r_0}}{(N_e+N_0)}}
\end{equation}

Which leads to the following slow-roll parameters:
\begin{eqnarray}
  \epsilon_1 &=& \frac{1}{2(N_e+N_0)}-\frac{C_2}{2{{(N_e+N_0)}^{2}}},\quad \epsilon_2={\frac{-(N_e+N_0)+2\,{C_2}}{(N_e+N_0) \left( -(N_e+N_0)+{C_2} \right) }} \\
  \delta_1&=& -{\frac {{r_0}-1}{N_e+N_0}}-{\frac {{C_2}}{{(N_e+N_0)}^{2}}},\quad \delta_2={\frac { \left( {r_0}-1 \right) (N_e+N_0)+2\,{C_2}}{(N_e+N_0) \left(  \left( {r_0}-1 \right) (N_e+N_0)+{C_2} \right) }}
\end{eqnarray}

 Solving equation $\epsilon_1(N_e=0)=1$, we obtain constant $C_2$:
\begin{equation}
 C_2=- \left( 2\,{ N_0}-1 \right) { N_0}\label{C2}
 \end{equation}

The substitution of \eqref{C2} to expressions for slow-roll parameters leads to:
\begin{eqnarray}
  \epsilon_2 &=& {\frac {4\,{{ N_0}}^{2}+N_e-{N_0}}{(N_e+N_0 )\left( 2\,{{N_0}}^{2}+N_e\right) }} ,\\
  \delta_1 &=& -{\frac {{r_0}\,(N_e+N_0)-2\,{{N_0}}^{2}-N_e}{(N_e+N_0)^{2}}}, \\
  \delta_2 &=&  {\frac { \left( {r_0}-1 \right) (N_e+N_0)-4\,{{ N_0}}^{2}+2\,{ N_0}}{(N_e+N_0) \left(  \left( {r_0}-1 \right) (N_e+N_0)-2\,{{N_0}}^{2}+{ N_0} \right) }}
\end{eqnarray}

At the end of inflation, this slow-roll parameters can be reduced to:
\begin{eqnarray}
\epsilon_2 &=&{\frac {4\,{ N_0}-1}{2{{N_0}}^{2}}} \\
  \delta_1 &=&  2-{\frac {{r_0}}{{ N_0}}} \label{d1}\\
   \delta_2 &=& {\frac {-{r_0}-1+4\,{ N_0}}{{ N_0}\, \left( -{r_0}+2\,{N_0} \right) }}
\end{eqnarray}

Evidently, to save the slow-roll regime we should have $-1<\delta_1<1$, considering the biggest $r_0$ and smallest $N_0$.  The~slow-roll parameter $\epsilon_2$
reaches the value $1$ at $N_e=0$ if $N_0=1+\frac{\sqrt{2}}{2}$ and after that, the slow-roll parameter $\epsilon_2$ grows and becomes bigger then $1$. Thus, we suppose that $N_0=1+\frac{\sqrt{2}}{2}$ is the smallest appropriate value for the sum $N_e+N_0$ during slow-roll~regime.

The value of constant $r_0$ included slow-roll parameters related to the restriction of the tensor-to-scalar ratio:
\begin{equation}
r=\frac{8r_0}{N_b+N_0}=0.065\cdot k, \quad \mbox{where} \quad 0<k<1, \quad N_b\quad\mbox{is a start point of inflation}
\end{equation}
and  parameter $r_0$ can be expressed as
\begin{equation}
r_0=\frac{0.065\cdot k\cdot (N_b+N_0)}{8}=0.008125\, k\, (N_b+N_0).\label{r0m}
\end{equation}

The  start point of inflation $N_b$ is related to the appropriate value of the spectral index:
\begin{equation}
n_s=1-\frac{2}{N_b+N_0}-\frac{(2N_0-1)N_0}{(N_b+N_0)^2}.\label{n_s}
\end{equation}

From here, we obtain an equation for $(N_b+N_0)$:
\begin{equation}
{\frac {-2\,{{N_0}}^{2}+{N_0}-2\,{(N_b+N_0)}}{{(N_b+N_0)}^{2}}}=n_s-1
\end{equation}
having two solutions:
\begin{eqnarray}
  N_b &=&{\frac {-1+\sqrt {-2\,{{N_0}}^{2}{n_s}+2\,{{ N_0}}^{2}+{ N_0}\,{n_s}-{ N_0}+1}}{{n_s}-1}}-N_0 \\
  N_b&=& -{\frac {1+\sqrt {-2\,{{N_0}}^{2}{n_s}+2\,{{ N_0}}^{2}+{ N_0}\,{n_s}-{ N_0}+1}}{{n_s}-1}}-N_0\label{Nb2}
\end{eqnarray}
but only the second solution allows to reproduce $55<N_b<65$.
To obtain $r_0$, we substitute \eqref{Nb2} into \eqref{r0m}.  {After that we} substitute the obtained $r_0$  into \eqref{d1}:
\begin{equation}
\delta_1=2+ \Delta\delta_2,\quad \Delta\delta_2={\frac {0.008125\,k \left( 1+\sqrt {-2\,{{N_0}}^{2}{n_s}+2\,{{N_0}}^{2}+{N_0}\,{n_s}-{N_0}+1} \right) }{
 \left( {n_s}-1 \right) {N_0}}}.
\end{equation}

To save slow-roll regime, the minimal condition  $-3<\Delta\delta_2<-1$ should be checked. Since $0<k<1$, we would like to analyze  $\Delta\delta_2/k$ using
the following inequality:
\begin{equation}
\frac{\Delta\delta_2}{k}=-{\frac {0.008125\, \left( 1+\sqrt {2\,{{N_0}}^{2}(1-{n_s})-{N_0}(1-{n_s})+1} \right) }{
 \left( 1-{n_s} \right) {N_0}}}<-1.
\end{equation}

From here, we can suppose:
\begin{equation}
{\frac {0.008125\, \left( 1+\sqrt {2\,{{N_0}}^{2}(1-{n_s})-{N_0}(1-{n_s})+1} \right) }{
 \left( {n_s}-1 \right) {N_0}}}=-l,\quad \mbox{where}\quad l>1.\label{***}
\end{equation}

The solution of \eqref{***} can be presented in the form:
\begin{equation}
n_s= 1-\frac{0.00013203125}{{l}^{2}}- \frac{0.01625}{l{N_0}}+ \frac{0.000066015625}{{l}^{2}{N_0}}.\label{*}
\end{equation}
which can be approximated:
\begin{equation}
n_s\approx1-\frac{0.01625}{l{N_0}}\quad
\end{equation}
due  to $l>1,$ $N_0>1$ and orders of numerical values of numbers included to \eqref{*}. From~here, it is evident that the increase in  $l$ and $N_0$ leads to the increase in $n_s$. We substitute the minimal values of parameters $l=1$, $N_0=1+\frac{\sqrt{2}}{2}$  to expression \eqref{*}  and obtain $n_s\approx0.99039$ as the minimal values of model spectral index. The~need value for the spectral index can be only be reached at $l<1$, leading to the breaking of the slow-roll~regime.

At the same time, the appropriate values of the spectral index lead to the deviation of a slow-roll regime via the parameter $\delta_1$.
Let us present this fact explicitly.
The  slow-roll parameter $\delta_1$ includes the constant $r_0$ which is related with the value of $N_b+N_0$.
The second solution \eqref{Nb2} to equation \eqref{n_s} can be presented in the form:
\begin{equation}
(N_b+N_0)=-{\frac {1+\sqrt {\left({N_0}-2\,{{N_0}}^{2}\right)(n_s-1)+1}}{n_s-1}}
\end{equation}

The substitution $N_0=1+\frac{\sqrt{2}}{2}$ to this equation leads to:
\begin{equation}
(N_b+N_0)=-{\frac {2+\sqrt {-(6\sqrt {2}+8)\,(n_s-1)+4}}{2(n_s-1)}}\label{Ns(z)}
\end{equation}

Using \eqref{Ns(z)}, the parameter $r_0$ is obtained, included in tensor-to-scalar ratio:
\begin{equation}
r_0=- {\frac {0.0040625 \,k \left( 2+\sqrt {-(6\,\sqrt {2}+8)\,(n_s-1)+4}
 \right) }{(n_s-1)}}
\end{equation}
and it is substituted for slow-roll parameter $\delta_1$:
\begin{equation}
\delta_1=2+{\frac {0.0040625 \,k \left( 2+\sqrt {-(6\,\sqrt {2}+8)\,(n_s-1)+4}
 \right) }{(n_s-1) \left( 1+1/\sqrt {2} \right) }}\label{delta1}
\end{equation}

The $(n_s-1) $ is less than zero and to obtain the smallest value of $\delta_1$, we assume $k=1$.
The consideration of \eqref{delta1} in the case of $k=1$ leads to $\delta_1>1$ at appropriate values of the spectral index. Let us present
minimal values of $\delta_1$ at key values of $n_s$:
\begin{enumerate}
  \item if $n_s=0.961$ then $\delta_1\geq1.7465$,
  \item if $n_s =0.965$ then $\delta_1\geq1.7186$,
  \item if $n_s =0.969$ then $\delta_1\geq1.6834$.
\end{enumerate}

The saving of appropriate values of spectral index $n_s=0.965\pm0.04$ leads to the divination of $\delta_1$ from the slow-roll regime during~inflation.

Thus, the~reconstruction of a minimally coupled model in EGB gravity leading to inflationary parameters
of the cosmological attractor with $r\sim(N_e+N_0)^{-1}$  during the slow-roll regime is~impossible.
\section{Conclusions}\label{sec4}
The introduction of effective potential and the tensor-to-scalar ratio allows to reproduce the
expression for $\xi$ in the case of minimal coupling between the field and the Ricci scalar.
In the case of nonminimal coupling, the~reproduction of $\xi$ will be related to the form of the coupling function.
The question of satisfaction of slow-roll
regime is rather important for analytically formulated models due to the effective potential approach
~\cite{Pozdeeva:2019agu,Pozdeeva:2021iwc} and to have the positive square of sound speed in Einstein--Gauss--Bonnet
gravity models~\cite{Pozdeeva:2020shl}. The~reheating after inflation is rather a popular problem~\cite{Bassett:2005xm,Martin:2021frd,DiMarco:2021xzk,Garcia:2020wiy}. However, the  reheating  is a rather open question in the Einstein--Gauss--Bonnet gravity~\cite{vandeBruck:2016xvt}.  { And the deviation from can coincide with start of prereheating processes}~\cite{Shojaee:2020xyr,Mohammadi:2020twg}. Therefore, we try to check satisfaction of the slow-roll regime during~inflation.

In the case of $r={8r_0}/{(N_e+N_0)^2}$ and  {the exponential} effective potential,
the slow-roll regime can be satisfied during inflation~\cite{Pozdeeva:2021iwc}.
However, the~slow-roll parameter $\delta_1$ related  {with} the e-folding number derivative of the function before the Gauss--Bonnet term
in the initial action $\xi^\prime$ leads to the deviation from the slow-roll regime for the models with  {the same}
effective potential and the following tensor-to-scalar ratio $r={8r_0}/{(N_e+N_0)}$. We demonstrate this using the representation of the function $\xi$ thought the effective potential and the tensor-to-scalar ratio. To~satisfy the restriction of observable data~\cite{Akrami:2018odb},
we restrict the value of parameter $r_0$  considering the tensor-to-scalar ratio $r$ in the beginning of inflation. In~the models with $r\sim(N_e+N_0)^{-2}$, the upper value of $r_0$ is bigger  then in models with  $r\sim(N_e+N_0)^{-1}$.
As a result, the value of $r\sim8r_0(N_e+N_0)^{-1}$ is rather small to satisfy the condition
\mbox{$|\delta_1|=(2\epsilon_1-r/8)<1$}  in the end of inflation $N_e=0$. We plan to generalize
the detailed analysis with the effective potential of the exponential~form to the case of nontrivial nonminimal coupling.

This work was partially supported by the Russian Foundation for Basic Research grant No.~20-02-00411.


\begin{thebibliography}{999}



\bibitem{Akrami:2018odb}
Akrami, Y.;~et~al. [{Planck Collaboration}
].Planck 2018 results. X. Constraints on inflation.
 \emph{ {Astron.Astrophys.}}%Please confirm journal name in all references. {\bf 2020},  {\em{641}}, A10.
[arXiv:1807.06211 [astro-ph.CO]]
%737 citations counted in INSPIRE as of 29 Apr 2020


\bibitem{Starobinsky:1980te}
 {Starobinsky, A.A.} %There are a lot of references with only one author name. Please check if more authors should be added.
 A New Type of Isotropic Cosmological Models Without Singularity.
    \emph{Phys.\ Lett. B}   {\bf 1980}, {\em 91}, 99--102.

\bibitem{Starobinsky:1982}
Starobinsky, A.A.
Dynamics of phase transition in the new inflationary universe scenario and generation of perturbations.
		\emph{\mbox{Phys. Lett. B}}   {\bf 1982}, {\em  117},  175--178.

\bibitem{Starobinsky:1983}
Starobinsky, A.A.
The Perturbation Spectrum Evolving from a Nonsingular Initially De-Sitter Cosmology and the Microwave Background Anisotropy.
 \emph{Sov. Astron. Lett.}  {\bf1983},  {\em 9}, {302--304}. %Please check all the single page numbers in the ref parts, if they are page range format, please complete the page numbers.


\bibitem{Mijic:1986iv}
  Mijic, M.B.; Morris, M.S.; Suen, W.M.
The $R^2$ Cosmology: Inflation Without a Phase Transition.
  \emph{Phys. Rev. D} {\bf 1986}, \emph{34}, 2934--2946.
 % doi:10.1103/PhysRevD.34.2934
  %%CITATION = doi:10.1103/PhysRevD.34.2934;%% 97

\bibitem{Maeda1988}
 Maeda, K.
 Inflation as a Transient Attractor in $R^2$ Cosmology.
 \emph{ Phys. Rev. D } {\bf 1988}, \emph{37},  858--862.
 % doi:10.1103/PhysRevD.37.858
  %%CITATION = doi:10.1103/PhysRevD.37.858;%% 96


\bibitem{Kallosh:2013hoa}
 Kallosh, R.; Linde, A.
 Universality Class in Conformal Inflation.
 \emph{ J. Cosmol. Astropart. Phys.} {\bf 2013},  { \emph{07}, 002.} [arXiv:1306.5220] %\emph{1307}, doi:10.1088/1475-7516/2013/07/002. 2. %MDPI: Please confirm the newly revised information
  %%CITATION = ARXIV:1306.5220;%% 51

%\cite{Galante:2014ifa}
\bibitem{Galante:2014ifa}
Galante, M.; Kallosh, R.; Linde, A.; Roest, D.
 Unity of Cosmological Inflation Attractors.
\emph{Phys. Rev. Lett.} {\bf 2015},  \emph{114}, 141302.
%doi:10.1103/PhysRevLett.114.141302
[arXiv:1412.3797]

  %\cite{Roest:2013fha}
\bibitem{Roest:2013fha}
Roest, D.
Universality classes of inflation.
		\emph{J. Cosmol. Astropart. Phys.} {\bf 2014},    {\emph{01}, 007}. 		[arXiv:1309.1285]


\bibitem{Barvinsky:1994hx}
 		Barvinsky, A.O.; Kamenshchik, A.Y.
Quantum scale of inflation and particle physics of the early universe. \emph{Phys. Lett. B} {\bf 1994},\mbox{ \emph{332}, 270--276.}[arXiv:gr-qc/9404062]

%\cite{Bezrukov:2007ep}
\bibitem{Bezrukov:2007ep}
  Bezrukov, F.L.; Shaposhnikov, M.
The Standard Model Higgs boson as the inflaton.
 \emph{Phys. Lett. B} {\bf 2008}, \emph{659}, 703--706.			
[arXiv:0710.3755]

%\cite{Barvinsky:2008ia}
\bibitem{Barvinsky:2008ia}
	Barvinsky, A.O.; Kamenshchik, A.Y.; Starobinsky, A.A.
Inflation scenario via the Standard Model Higgs boson and LHC.			 \emph{J. Cosmol. Asropart. Phys.} \textbf{2008},   {\emph{811}, 21}.
[arXiv:0809.2104]

%\cite{Bezrukov:2010jz}
\bibitem{Bezrukov:2010jz}
Bezrukov, F.L.; Magnin, A.; Shaposhnikov, M.; Sibiryakov, S.
Higgs inflation: Consistency and generalizations.
\emph{J. High Energy Phys.} {\bf 2011},  {\emph{01},  016.}
[arXiv:1008.5157]	


%\cite{Bezrukov:2013fka}
\bibitem{Bezrukov:2013fka}
Bezrukov, F.L.
The Higgs field as an inflaton.
\emph{Class. Quant. Grav.} {\bf 2013}, \emph{30},  214001.
	[arXiv:1307.0708]

%\cite{Rubio:2018ogq}
\bibitem{Rubio:2018ogq}
Rubio, J.
 Higgs inflation.
\emph{Front. Astron. Space Sci.} {\bf 2019}, \emph{5}, 50.
[arXiv:1807.02376]


\bibitem{Cervantes-Cota1995}
   		 Cervantes-Cota, J.L.; Dehnen,  H.
		 Induced gravity inflation in the standard model of particle physics.
		 \emph{Nucl.\ Phys. B}~{\bf 1995}, \emph{442}, 391--412. [arXiv:astro-ph/9505069]

%\cite{Elizalde:2014xva}
\bibitem{Elizalde:2014xva}
Elizalde, E.;  Odintsov, S.D.; Pozdeeva, E.O.; Vernov, S.Y.
Renormalization-group inflationary scalar electrodynamics
and $SU(5)$ scenarios confronted with Planck2013 and BICEP2 results. {\em  Phys.\ Rev. D}  \textbf{2014},  \emph{90,} 084001. [arXiv:1408.1285 [hep-th]]

%41 citations counted in INSPIRE as of 06 May 2021

\bibitem{Elizalde:2015nya}
Elizalde, E.;  Odintsov, S.D.; Pozdeeva, E.O.; Vernov, S.Y.
Cosmological attractor inflation from the RG-improved Higgs sector of finite gauge theory. \emph{J. Cosmol. Astropart. Phys.} \textbf{2016}, \emph{1602}, 25.  [arXiv:1509.08817]


 %\cite{Kaiser:2012ak}
\bibitem{Kaiser:2012ak}
Kaiser, D.I.; Mazenc, E.A.; Sfakianakis, E.I.
Primordial Bispectrum from Multifield Inflation with Nonminimal Couplings.
 \emph{Phys. Rev. D} \textbf{2013}, \emph{87}, 064004.
[arXiv:1210.7487]


 	
%\cite{Greenwood:2012aj}
\bibitem{Greenwood:2012aj}
Greenwood, R.N.; Kaiser, D.I.; Sfakianakis, E.I.
		Multifield Dynamics of Higgs Inflation.
		 \emph{Phys. Rev. D} \textbf{2013}, \emph{87}, 064021.
	[arXiv:1210.8190]



\bibitem{Dubinin:2017irg}
Dubinin,  M.N.; Petrova, E.Y.; Pozdeeva, E.O.; Sumin, M.V.; Vernov, S.Y.
MSSM-inspired multifield inflation.
 \emph{J. High Energy Phys.}  \textbf{2017}, \emph{1712}, 36. [arXiv:1705.09624]



%\cite{Dubinin:2017irq}
\bibitem{Dubinin:2017irq}
Dubinin,  M.N.; Petrova, E.Y.; Pozdeeva, E.O.; Vernov, S.Y.
MSSM inflation and cosmological attractors.
\emph{Int. J. Geom. Meth. Mod. Phys.} {\bf 2018},  \emph{15}, 1840001.
%doi:10.1142/S0219887818400017
[arXiv:1712.03072]

%
\bibitem{Antoniadis:1993jc}
  Antoniadis,  I.; Rizos, J.; Tamvakis, K.
  {Singularity-free cosmological solutions of the superstring effective action.}
  \emph{Nucl. Phys. B} \textbf{1994}, \emph{415},  497--514.
 [arXiv:hep-th/9305025]


\bibitem{Torii:1996yi}
 Torii,  T.; Yajima, H.; Maeda, K.I.
  {Dilatonic black holes with Gauss-Bonnet term}.
  \emph{Phys.\ Rev. D} {\bf 1997}, \emph{55}, 739--753.
   [arXiv:gr-qc/9606034]


\bibitem{Kawai1998}
  Kawai,  S.; Sakagami, M.A.; Soda, J.
  {Instability of one loop superstring cosmology.}
 \emph{ Phys. Lett. B} {\bf 1998}, \emph{437}, 284--290 , [arXiv:gr-qc/9802033]

%\cite{Calcagni:2005im}
\bibitem{Calcagni:2005im}
Calcagni, G.; Tsujikawa, S.; Sami,  M.
 Dark energy and cosmological solutions in second-order string gravity.  \emph{Class. Quant. Grav.} \textbf{2005}, \emph{22}, 3977--4006.
 [arXiv:hep-th/0505193]

%\cite{Cartier:2001is}

\bibitem{Cartier:2001is}
Calcagni, G.; Hwang, J.C.; Copeland, E.J.
Evolution of cosmological perturbations in nonsingular string cosmologies.
  \emph{\mbox{Phys. Rev. D}}  {\bf 2001}, \emph{64}, 103504.
  [astro-ph/0106197]

  %\cite{Hwang:2005hb}
\bibitem{Hwang:2005hb}
   Hwang, J.C.; Noh,  H.
   Classical evolution and quantum generation in generalized gravity theories including string corrections and tachyon: Unified analyses.
  \emph{Phys. Rev. D} \textbf{2005}, \emph{71}, 063536.
     [arXiv:gr-qc/0412126]


\bibitem{Tsujikawa:2006ph}
  Tsujikawa,  S.; Sami,  M.
String-inspired cosmology: Late time transition from scaling matter era to dark energy universe caused by a Gauss-Bonnet coupling.  \emph{ J. Cosmol. Astropart. Phys.} \textbf{2007},  {\emph{01},  006.}   [arXiv:hep-th/0608178]




\bibitem{Cognola:2006sp}
  Cognola, G.; Elizalde, E.;  Nojiri, S.; Odintsov, S.D.; Zerbini, S.
  { String-Inspired Gauss-Bonnet gravity reconstructed from the universe expansion history and yielding the transition from matter dominance to dark energy}.
 {\em  Phys. Rev. D} \textbf{2007}, \emph{75}, 086002.
 [arXiv:hep-th/0611198]


\bibitem{Pozdeeva:2020apf}
 {Pozdeeva, E.O.; Gangopadhyay,  M.R.; Sami, M.; Toporensky, A.V.; Vernov, S.Y.
Inflation with a quartic potential in the framework of Einstein-Gauss-Bonnet gravity.} %References 31 and 58 are the same. Please revise.
\emph{Phys. Rev. D} \textbf{2020}, \emph{102}, 043525.  [arXiv:2006.08027]


\bibitem{Soda2008}
Satoh, M.; Soda, J.
 Higher Curvature Corrections to Primordial Fluctuations in Slow-roll Inflation.
\emph{J. Cosmol. Astropart. Phys.} {\bf 2008}, \emph{9}, 19.
[arXiv:0806.4594]


\bibitem{Guo:2009uk}
 Guo, Z.K.; Schwarz,  D.J.
   Power spectra from an inflaton coupled to the Gauss-Bonnet term.
  \emph{Phys. Rev. D} {\bf 2009}, \emph{80}, 063523.
[arXiv:0907.0427]

%\cite{Guo:2010jr}
\bibitem{Guo:2010jr}
Guo, Z.K.; Schwarz,  D.J.
Slow-roll inflation with a Gauss-Bonnet correction.
{\em Phys. Rev. D} \textbf{2010}, \emph{81},  123520.
 [arXiv:1001.1897]

\bibitem{Jiang:2013gza}
Jiang,  P.X.; Hu, J.W.; Guo, Z.K.
   Inflation coupled to a Gauss-Bonnet term.
  \emph{Phys.\ Rev. D} \textbf{2013}, \emph{88}, 123508.
  [arXiv:1310.5579]

%\cite{Koh:2014bka}
\bibitem{Koh:2014bka}
Koh, S.; Lee, B.H.; Lee, W.; Tumurtushaa, G.
{Observational constraints on slow-roll inflation coupled to a Gauss-Bonnet term}. \emph{ Phys. Rev. D} \textbf{2014}, \emph{90},  063527.
[arXiv:1404.6096]

%\cite{Koh:2016abf}
\bibitem{Koh:2016abf}
Koh, S.; Lee, B.H.; Tumurtushaa, G.
Reconstruction of the Scalar Field Potential in Inflationary Models with a Gauss-Bonnet term.
\emph{Phys. Rev. D} \textbf{2017}, \emph{95},  123509.
[arXiv:1610.04360]

%\cite{Koh:2018qcy}
\bibitem{Koh:2018qcy}
Koh, S.; Lee, B.H.; Tumurtushaa, G.
Constraints on the reheating parameters after Gauss-Bonnet inflation from primordial gravitational waves.
\emph{Phys. Rev. D} \textbf{2018}, \emph{98}, 103511.
[arXiv:1807.04424]

\bibitem{vandeBruck:2015gjd}
  van de Bruck, C.; Longden, C.
   {Higgs Inflation with a Gauss-Bonnet term in the Jordan Frame}.
  {\em  Phys. Rev. D} \textbf{2016}, \emph{93},  063519.
   [arXiv:1512.04768]

\bibitem{JoseMathew}
Mathew, J.; Shankaranarayanan, S.
Low scale Higgs inflation with Gauss-Bonnet coupling.
\emph{Astropart. Phys}. {\textbf{2016}},  {\emph{84}, 001.}   [arXiv:1602.00411]


\bibitem{vandeBruck:2016xvt}
 van de Bruck, C.; Longden, C.; Dimopoulos, K.
 Reheating in Gauss-Bonnet-coupled inflation.
  \emph{Phys. Rev. D} \textbf{2016}, {\emph{94}}, 023506.
[arXiv:1605.06350]


\bibitem{Nozari:2017rta}
  Nozari  K.; Rashidi, N.
   {Perturbation, Non-Gaussianity, and reheating in a Gauss-Bonnet $\alpha$-attractor model}.
 \emph{ Phys. Rev. D} \textbf{2017}, \mbox{\emph{95}, 123518.}
  [arXiv:1705.02617]

\bibitem{Armaleo:2017lgr}
Armaleo, J.M.; Osorio Morales, J.; Santillan, O.
Gauss-Bonnet models with cosmological constant and non zero spatial curvature in $D=4$.
\emph{Eur. Phys. J. C}  \textbf{2018}, \emph{78}, 85. [arXiv:1711.09484]

\bibitem{Chakraborty:2018scm}
Chakraborty, S.; Paul T.; SenGupta, S.
 {Inflation driven by Einstein-Gauss-Bonnet gravity}.
\emph{Phys. Rev. D} \textbf{2018}, \emph{98}, 083539.
[arXiv:1804.03004]

\bibitem{Yi:2018gse}
Yi, Z.; Gong, Y.; Sabir, M.
Inflation with Gauss-Bonnet coupling.
\emph{Phys. Rev. D} \textbf{2018}, \emph{98}, 083521.[arXiv:1804.09116]

\bibitem{Odintsov:2018zhw}
Odintsov, S.D.; Oikonomou,  V.K.
Viable Inflation in Scalar-Gauss-Bonnet Gravity and Reconstruction from Observational Indices.
  \emph{Phys. Rev. D} \textbf{2018}, \emph{98}, 044039.
[arXiv:1808.05045]


%\cite{Nojiri:2019dwl}
\bibitem{Nojiri:2019dwl}
Nojiri, S.;  Odintsov, S.D.; Oikonomou, V.K.; Chatzarakis, N.; Paul, T. Viable inflationary models in a ghost-free Gauss\textendash{}Bonnet theory of gravity.
\emph{Eur. Phys. J. C} \textbf{2019}, \emph{79},  565.
[arXiv:1907.00403 [gr-qc]]

\bibitem{Kleidis:2019ywv}
Kleidis, K.; Oikonomou, V.
A Study of an Einstein Gauss-Bonnet Quintessential Inflationary Model.
\emph{Nucl. Phys. B} \textbf{2019}, \mbox{\emph{948}, 114765.}
[arXiv:1909.05318]

\bibitem{Rashidi:2020wwg}
Rashidi, N.; Nozari  K.
 Gauss-Bonnet Inflation after Planck2018.
\emph{Astrophys. J.} \textbf{2020}, \emph{890}, 58.
[arXiv:2001.07012]

\bibitem{Odintsov:2020sqy}
Odintsov, S.D.; Oikonomou, V.K.; Fronimos, F.P.
Rectifying Einstein-Gauss-Bonnet Inflation in View of GW170817.
\emph{Nucl. Phys. B} \textbf{2020}, \emph{958}, 115135.
[arXiv:2003.13724]

%\cite{Odintsov:2020zkl}
\bibitem{Odintsov:2020zkl}
Odintsov, S.D.; Oikonomou,  V.K.
Swampland Implications of GW170817-compatible Einstein-Gauss-Bonnet Gravity.
 \emph{Phys. Lett.  B} \textbf{2020}, \emph{805}, 135437.
 [arXiv:2004.00479]


\bibitem{Hikmawan:2015rze}
Hikmawan, G.; Soda, J.; Suroso, A.; Zen, F.P.
 Comment on ``Gauss-Bonnet inflation''.
 \emph{Phys. Rev. D} \textbf{2016}, \emph{93}, 068301.
[arXiv:1512.00222]



\bibitem{Mukhanov:2013tua}
Mukhanov, V.
Quantum Cosmological Perturbations: Predictions and Observations.
\emph{Eur. Phys. J. C}  \textbf{2013}, \emph{73}, 2486.
[arXiv:1303.3925]


%\cite{Pozdeeva:2020shl}
\bibitem{Pozdeeva:2020shl}
Pozdeeva, E.O.
Generalization of cosmological attractor approach to Einstein\textendash{}Gauss\textendash{}Bonnet gravity.
\emph{Eur. Phys. J. C} \textbf{2020}, \emph{80},  612.
[arXiv:2005.10133 [gr-qc]]



\bibitem{Pozdeeva:2021iwc}
Pozdeeva, E.O.;  Vernov, S.Y.
Construction of inflationary scenarios with the Gauss-Bonnet term and nonminimal coupling. [arXiv:2104.04995 [gr-qc]]
%1 citations counted in INSPIRE as of 01 May 2021 81

\bibitem{Leach:2002ar}
Leach, S.M.; Liddle, A.R.; Martin, J.; Schwarz, D.J.
Cosmological parameter estimation and the inflationary cosmology.
\emph{\mbox{Phys. Rev. D}} \textbf{2002}, \emph{66}, 023515.
[astro-ph/0202094 [astro-ph]]

\bibitem{Linde:1983gd}
Linde, A.D.
Chaotic Inflation.
\emph{Phys. Lett. B } \textbf{1983}, \emph{129}, 177.
%doi:10.1016/0370-2693(83)90837-7
%3064 citations counted in INSPIRE as of 07 May 2021

\bibitem{Pozdeeva:2019agu}
 {Pozdeeva, E.O.;  Sami, M.; Toporensky, A.V.; Vernov, S.Y. Stability analysis of de Sitter solutions in models with the Gauss-Bonnet term.}
{\em Phys. Rev. D} \textbf{2019}, \emph{100}, 083527.
[arXiv:1905.05085]

  %\cite{Vernov:2021hxo}
\bibitem{Vernov:2021hxo}
Vernov, S.Y.; Pozdeeva, E.O.
De Sitter solutions in Einstein-Gauss-Bonnet gravity.
\emph{Universe} \textbf{2021}, \emph{7},  149.
[arXiv:2104.11111 [gr-qc]]


%\cite{Bassett:2005xm}
\bibitem{Bassett:2005xm}
Bassett, B.A.; Tsujikawa, S.; Wands,  D.
Inflation dynamics and reheating.
\emph{Rev. Mod. Phys.} \textbf{2006},  \emph{78}, 537--589.
[arXiv:astro-ph/0507632]
%803 citations counted in INSPIRE as of 25 May 2021

%\cite{Martin:2021frd}
\bibitem{Martin:2021frd}
Martin, J.; Pinol, L.
Opening the reheating box in multifield inflation.
[arXiv:2105.03301]
%0 citations counted in INSPIRE as of 25 May 2021

%\cite{DiMarco:2021xzk}
\bibitem{DiMarco:2021xzk}
Marco, A.D.;  Pradisi, G.
Variable Inflaton Equation of State and Reheating.  [arXiv:2102.00326]
%3 citations counted in INSPIRE as of 25 May 2021


%\cite{Garcia:2020wiy}
\bibitem{Garcia:2020wiy}
Garcia, M.A.G.; Kaneta, K.; Mambrini,Y.; Olive,  K.A.
Inflaton Oscillations and Post-Inflationary Reheating.
\emph{JCAP} \textbf{2021}, \emph{4}, 12.
%doi:10.1088/1475-7516/2021/04/012
[arXiv:2012.10756]
%5 citations counted in INSPIRE as of 25 May 2021

%\cite{Shojaee:2020xyr}
\bibitem{Shojaee:2020xyr}
Shojaee, R.; Nozari, K.; Darabi, F.
\ensuremath{\alpha}-Attractors and reheating in a nonminimal inflationary model.
\emph{Int. J. Mod. Phys. D} \textbf{2020}, \mbox{\emph{29}, 2050077.}
[arXiv:2101.03981]
%2 citations counted in INSPIRE as of 25 May 2021

%\cite{Mohammadi:2020twg}
\bibitem{Mohammadi:2020twg}
Mohammadi, A.; Golanbari, T.; Enayati, J.; Jalalzadeh, S.; Saaidi,  K.
Revisiting Scalar Tensor inflation by swampland criteria and Reheating.  [arXiv:2011.13957]
%3 citations counted in INSPIRE as of 25 May 2021

\end{thebibliography}
\end{document}